\documentclass[conference,10 point]{IEEEtran}

\usepackage[boxed,algosection,linesnumbered]{algorithm2e}
\usepackage{epsfig}
\usepackage{bbm}
\usepackage{fullpage}
\usepackage{amssymb}
\usepackage{amsmath}
\usepackage{amscd}
\usepackage{amsthm}
\usepackage{multirow}

\begin{document}
%
\title{CloudSimNFV: Modeling and Simulation of Energy-Efficient NFV in Cloud Data Centers}

\author{\IEEEauthorblockN{Wutong Yang, Minxian Xu, Guozhong Li}
\IEEEauthorblockA{University of Electronic Science and\\
Technology of China\\
Email: uestcywt@foxmail.com,\\
xmxyt900@gmail.com,\\
GuozhongLi@hotmail.com}\\
\and
\IEEEauthorblockN{Wenhong Tian}
\IEEEauthorblockA{University of Electronic Science and Technology of China, \\
BigData Research Center, UESTC, China \\
Email: tian\_wenhong@uestc.edu.cn, }

}

\maketitle
\begin{abstract}
Network Function Virtualization (NFV) takes advantage of hardware virtualization to undertake software processing for various functions, and complements the drawbacks of traditional network technology. To speed up NFV related research, we need a user friendly and easy to use research tool, which could support data center simulation, scheduling algorithms implementation and extension, and provide energy consumption simulation. As a cloud simulation toolkit, CloudSim has strong extendibility that could be extended to simulate NFV environment. This paper introduces a NFV cloud framework based on CloudSim and an energy consumption model based on multi-dimensional extension, implementing a toolkit named ClousimNFV to simulate the NFV scenario, proposing several scheduling algorithm based on for NFV applications. The toolkit validation and algorithm performance comparison are also given.
\end{abstract}

\IEEEpeerreviewmaketitle

\section{Introduction}
In traditional data centers, applications are tied to specific physical servers that are often over-provisioned to deal with upper-bound workload. Such configuration makes data centers expensive to maintain with wasted energy and floor space, low resource utilization and significant management overhead. With virtualization technology, today¡¯s Cloud data centers become more flexible, secure and provide better support for on-demand allocating \cite{Buyya}.

Currently, computation or storage is treated as independent of network, and energy efficiency is not considered enough, which leads the service values are not maximized. The motivations of Network Function Virtualization (NFV) comes to satisfy the aim that putting the network functions in Clouds, combining values (Infrastructure as a service, Compute/Storage, Network infrastructure as a service, etc.), leveraging NFV infrastructure of other service providers to increase resiliency, reducing latency and regulatory requirements \cite{Chiosi}. Moreover, the ultimate goal for NFV is transforming network architectures by implementing Network Functions in Software that can run on commodity hardware \cite{Udupi}.

To further foster innovation and development, we require toolkits that provide a testbed for experiments with Network Function Virtualization systems within a cloud data center. While there isn't any tool that can fulfill this goal yet. \textbf{To accelerate the related research on NFV, we introduce CloudSimNFV that enables to simulate scenario for NFV.}

CloudSimNFV is a newly developed tool built on top of CloudSim \cite{IEEEhowto:Calheiros}. In this paper, we discuss the innovation of CloudSimNFV and introduce its detailed design implementation. A framework is designed to evaluate resource management policies applicable for NFV scenario under cloud data center environment. It enables to simulate cloud data centers, physical machines, virtual machines and NFV applications. CloudSimNFV can also implement energy consumption model with predefined loads to monitor the energy cost in data centers. Moreover, CloudSimNFV provides portal pages to simplify the simulation configurations and show the simulation results.

CloudSimNFV validation is tested with realistic communication service providers' data. The loads coming into data centers are periodically fluctuating over time referring to realistic communication providers data. CloudSimNFV has implemented several scheduling algorithms for resource allocation to compare performance under several scenarios. The implemented algorithms have their specific advantages and are possible to be applied to different NFV scenarios.

 The remainder of this paper is as follows. In Section II, we introduce the related work on NFV and why we choose CloudSim to extend NFV scenario. In Section III, we present the requirements for a NFV simulation tool combining the features of NFV scenarios and applications. The CloudSimNFV framework and energy consumption model are demonstrated in Section IV. The experiments and performance comparison are given in Section V. Finally, a conclusion and future work are concluded in Section VI.

\section{Related Work}
There have been some research improved the related research on NFV. Basta et al. \cite{Basta} proposed the functions placement problem in practice, and predicted the way that applying NFV and SDN to LTE mobile gateways to solve the problem. Carella et al. \cite{Carella} introduced an IP multi-media subsystem based on Network Function Virtualization architecture. Bolla et al \cite{IEEEhowto:Bolla} presented an enhanced framework named DROPv2, which contains power management mechanism to satisfy the energy efficiency of Network Function Virtualization. Raggio et al. \cite{Riggio} introduced EmPOWER as a novel and open platform for future tests and research on Network Function Virtualization. Udupi et al. \cite{Udupi} emphasized the combination of NFV and Openstack platform that a smart scheduler component could help improve the efficiency of infrastructure. Apart from these mentioned work, an easy to use and repeatable toolkit for NFV is still needed.

Quite a few cloud simulation tools have been implemented to simulate cloud data centers and compare resource scheduling algorithms. A survey paper summarized by Tian et al \cite{Tian1} has compared and discussed several simulators, like CloudSim \cite{IEEEhowto:Calheiros}, GreenCloud \cite{IEEEhowto: Kliazovich}, iCanCloud \cite{Nunez}, CloudSched \cite{IEEEhowto:Tian2} from their architecture view, modeling view, performance evaluation view and so on. Xu \cite{Xu} proposed a flexible and light-weight simulator, namely FlexCloud, for testing cloud resource scheduling algorithms. All these simulators have their strengths and weaknesses. Considering extendibility, implemented components (physical machines, virtual machines and task loads) and adopted energy consumption model. CloudSim is the easiest choice to extend and implement the NFV feature.

\section{Network Function Virtualization for Cloud Data Centers} \label{Model}
Simulation system has huge advantage for researching migration policies and energy consumption evaluation in NFV environment, as it need not to be built on complex real environment and can be sustainable to process simulation tests for a long period. It could also run experiments quickly and repeatedly, and provide detailed results for analysis. As for our algorithm evaluation, we can regard the algorithm as the initial filter that could pick up the suitable resource. In addition, with simulation tools, as the simulation time is quite less than realistic tests, much more test cases could be tested for reference. When we are comparing the efficiency of algorithms, we could obtain an approximate result for expectation before we undertake real tests. Therefore, our main goals focus on repeatability and controllability. We identify the following requirements:\\
1. Enable to simulate the data center core computing elements;\\
2. Enable to simulate variable work loads, which could fluctuate in real time;\\
3. Enable to simulate different migration policies;\\
4. Enable to simulate different energy consumption models, which could switch to different model and evaluate energy consumption.\\
5. Enable to simulate Network Function Virtualization applications.

\section{CloudSimNFV Framework Design}
\subsection{CloudSimNFV Architecture}
Our NFV simulation tool, CloudSimNFV is based on CloudSim platform. Taking advantage of data center simulation of CloudSim, we extend the specific NFV scenario that we require. We add the energy consumption model and extend scheduling algorithms under NFV scenario. As for the dynamic loads, we can simulate through loading the load files. Program codes (which are written in Java) can provide physical and virtual topology configurations. Another approach for user input is a portal that translates requirements into physical and virtual topology configurations.

Fig.1 shows the overall architecture of CloudSimNFV. Firstly, the user code and scenarios are composed of two modules: simulation specification and scheduling policy, which is configured by users and defined by developers. The simulation specification module defines the NFV scenario that simulation would be executed in, user requirements and energy aware characteristic. Scheduling policy module includes implemented and to be extended scheduling policies, which can be divided as VM migration policy, VM selection, initial allocation policy according to their different scheduling stages.

\begin{figure*}[!ht]
\begin{center}
{\includegraphics [width=1.0\textwidth, height=4in,angle=-0] {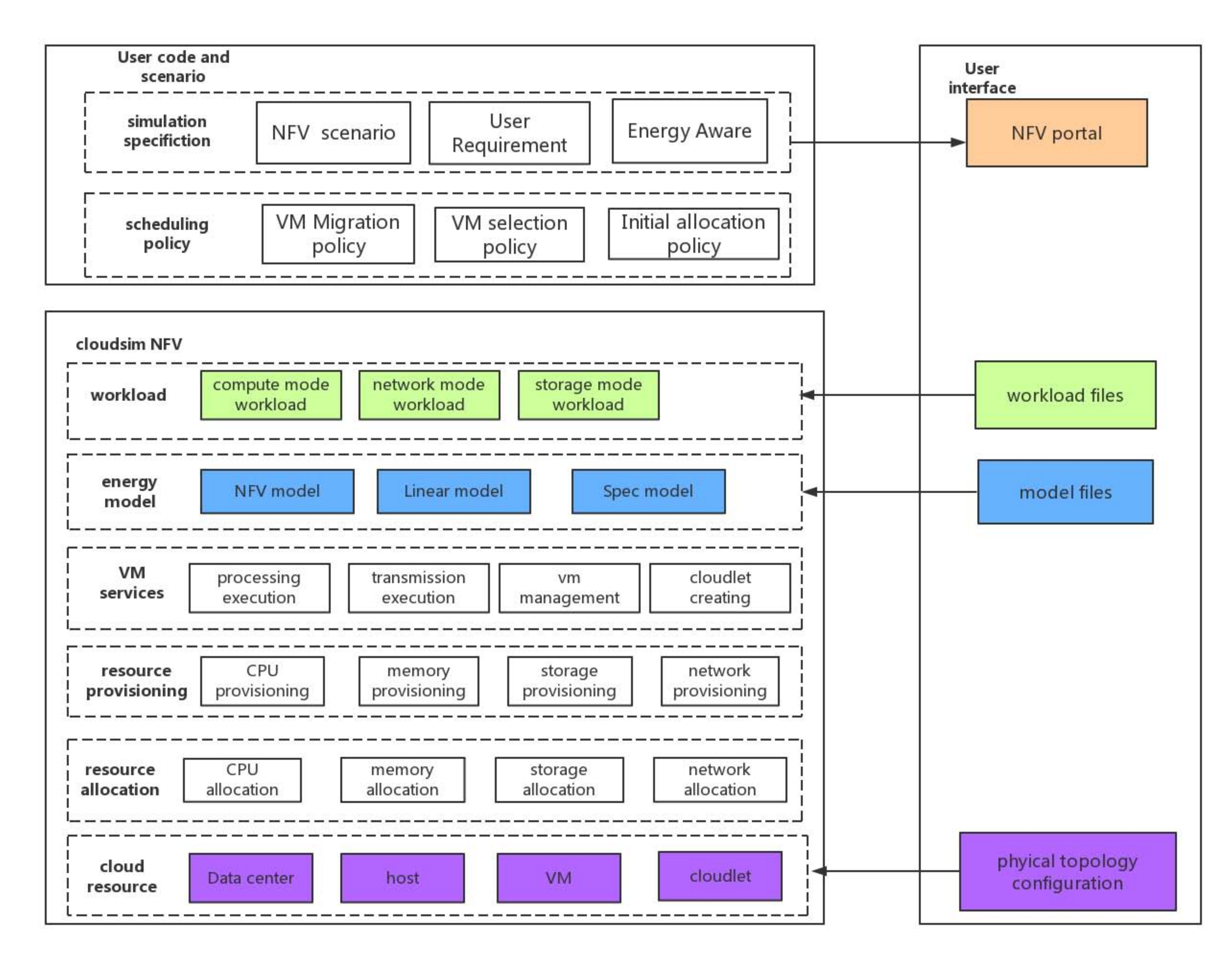}}
\caption{CloudSimNFV Architecure}
\end{center}
\end{figure*}

Besides the description of the user code and scenario, the end-users' requests, which composes the input workload for the simulation and is supplied in TXT files. Each type of the workload includes four workload files which are used to simulate cpu workload, memory workload, storage workload and bandwidth workload separately. These workload files can form different modes of workloads, like compute mode workload (CPU resource is required most), network mode workload (bandwidth resource is required most), storage model workload (storage resource is required most). Different types of workload can simulate diversified NFV working environment, which help us to test the energy consumption of each situation.

The lower layer is energy model layer. Users can specify the energy model according to the actual scenario (We will introduce an energy model in Section IV.B). Corresponding to workload and energy model, the provided scheduling policies, such as VM migration policy, VM selection policy and initial allocation policy are supported by the user code and scenario module. Brokers can be programmed to simulate the behavior of end-users or data centers. Regarding these policies, a user can either utilize built-in policies or can develop their own (by extending abstract classes).

VM Services are in charge of managing VMs, by calculating application execution and packet transmission time between VMs, it also supplies the creation of cloudlet and the updating of VM state. The next layer, Resource Provisioning, is composed of four modules. CPU Provisioning is a module to provide CPU resource to VM, memory Provisioning provides memory resource to VM, disk Provisioning provides storage resource to VM, and network Provisioning is provision of network resource to VM. Resource Provisioning depends on the workload that end-user chooses to use. The next layer, Resource Allocation, containing modules that allocate resources in the bottom layer of the architecture, according to the VM migration policy and VM selection policy specified by simulator users. These three layers are adopted from the original CloudSim codes.

The bottom layer of this architecture is the cloud resource layer. This layer contains the data center, host, VM and cloudlet components to simulate the different components of real data centers. Some extension work has been done in this layer. We extend the cloudlet as NFVlet to simulate the NFV task under NFV scenario (more details are given in IV.C). Physical topology configuration is implemented to represent the relationship between these components and can be configured in user interface.

\subsection{Energy Consumption Model}

The system energy consumption model is modeling for the whole system. Many researches have found that CPU utilization is typically proportional to the overall system load, and with realistic tests, a power consumption model for blade server is proposed as below \cite{Economou}:
\begin{equation}
\scriptsize
P=14.5+0.2U_{cpu} + (4.5e^{-8})U_{mem} + 0.003U_{disk} + (3.1e^{-8})U_{net}
\end{equation}
Where $U_{cpu}, U_{mem}, U_{disk}, U_{net}$ are utilization of CPU, memory, storage and network respectively. It can be seen that other factors have very small impact on total energy consumption. The parameter value could be tested and determined with specific server, such as combining learning and manual intervention. The parameters may be different for different servers.

When virtual machine is allocated to physical server, given that the CPU capacity occupied by virtual machine is just the capacity submitted by virtual machine, then the increased CPU utilization of physical server after virtual machine allocation can be calculated as:
\begin{equation}
\triangle u = \frac{VM.cpu}{PM.cpu}
\end{equation}
where $VM.cpu$ is the CPU capacity of virtual machine, $PM.cpu$ is the CPU capacity of physical machine. Then the VM energy consumption can be calculated as:
\begin{equation}
E_{vm} = (P_{max} - P_{min})\times (t_1 - t_0) \times \frac{VM.cpu}{PM.cpu}
\end{equation}
Energy consumption of physical server could be calculated as the turned on energy consumption (when CPU utilization is 0) adds the virtual machine energy consumption running on the machine. $E_{poweron}$ is the power when server is turned on, $E_{pm}$ is the server energy consumption:
\begin{equation}
E_{poweron} = P_{min} \times T_{poweron}
\end{equation}
\begin{equation}
E_{pm} = E_{poweron} + \sum^{n}_{i=1}{E_{vmi}}
\end{equation}
where $T_{poweron}$ is the running time of physical server, $E_{vmi}$ is the energy consumption of the $i-th$ virtual machine, $n$ is the virtual machine number on physical server. The total energy consumption of data center:
\begin{equation}
E_{DC} = \sum^{n}_{i=1}{E_{pmi}}
\end{equation}
where $E_{DC}$ is the total energy consumption of all physical servers, $E_{pmi}$ is the energy consumption of the $i-th$ physical machine, $n$ is the sum of all physical servers.

Fig. 2 shows the energy consumption calculation implementation structure. The structure is mainly composed of the following components:

\emph{CloudSim core logic:} The original CloudSim core logic is used to simulate the basic compute elements that compose the cloud infrastructure. On CloudSim, physical hosts can be defined with specific configurations and VMs are placed on the host that meets resource requirements such as CPU power, memory, network bandwidth and storage size. CloudSim simulates a range of elements of the cloud architecture, including data center, physical host, VM, VM scheduler, workload scheduler, etc.

\emph{VM scheduler modules}: \emph{VmScheduler} policy of NFV not only considers the CPU resource, but also takes all elements into account. VMs are placed in physical hosts according to \emph{VmAllocation} policy. During the period of simulation, hosts migrate VMs using the VM scheduler policy specified by users.And \emph{VmScheduler} policy would combine cpu resource ,ram resource, disk resource and network resource to decide which VM should be migrated to a new place. VM owns one or many cloudlets, and each cloudlet is described with the required computational power, memory, and storage size. Once the cloudlet is placed in certain VM, it will exist until cloudlet's life cycle is over.

\emph{Energy aware modules: }In order to evaluate the energy consumption of NFV environment, we developed a \emph{PowerModelNFV} class which makes our energy consumption calculation more precisely. It evaluates weighted cpu utilization, ram utilization, disk utilization and network utilization. All the resources of \emph{PowerHost} are provisioned by \emph{RamProvisioner}, \emph{BwProvisioner} and \emph{DiskProvisioner} .
There are three VM allocation policies we have appended, these VM allocation policies are inspired by NFV algorithm, ecoCloud algorithm and DRS algorithm. And all the VM allocation policies we developed need to extend the \emph{PowerVmAllocationPolicyMigrationAbstract} class.

\begin{figure}[!ht]
\begin{center}
{\includegraphics [width=0.5\textwidth, height=2.5in,angle=-0] {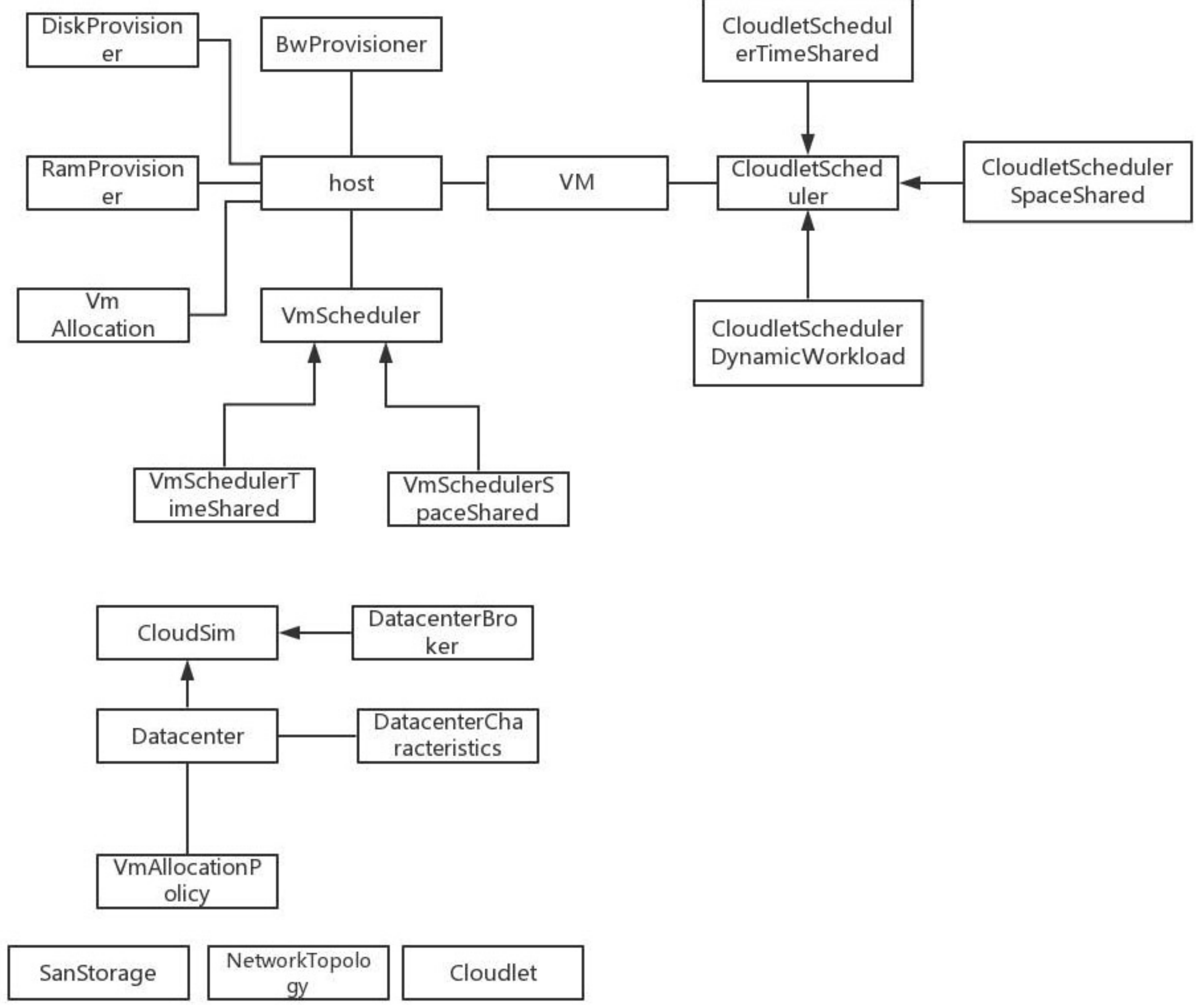}}
\caption{CloudSimNFV Energy Consumption Model Implementation Structure}
\end{center}
\end{figure}

\subsection{Load Generation}
Energy consumption tests under simulation environment are built on data center simulator CloudSim. In the CloudSim version after 3.0, Dynamic Voltage Frequency Scale (DVFS)\cite{Guerout} technique has been added into CloudSim. With the combination of CloudSim + DVFS, the model elements include physical machine (PM), virtual machine (VM) and cloudlet. Cloudlet is regarded as the tasks coming into system, which in our model, we transform cloudlet as NFV application as NFVlet. The physical machine modeling and virtual machine modeling are as same as the modeling in CloudSim.

We implement the aforementioned energy consumption model in previous section, then define the resource that PM may provide, VM and NFVlet may require. An simple example is presented below:\\
$PM: 1000MIPS$\\
$VM1: 100MIPS£¬NFVlet1: 150MI$\\
$VM2: 20MIPS£¬NFVlet2: 80MI$

Considering this example, with composition of different VM and NFVlet, various types of CPU loads could be generated. Such as $C1=[VM1, NFVlet1]$   $C2=[VM1, NFVlet2]$ $C3=[VM2, NFVlet1]$   $C4=[VM2, NFVlet2]$

$C1$ could produce 1.5s duration 10\% CPU load, $C2$ could produce 0.8s duration 10\% CPU load, $C3$ could produce 7.5s 2\% CPU load, $C4$ could produce 4s duration 2\% CPU load. Besides that, more various compositions could produce more types of loads with various durations and CPU loads, simulating resources and tasks under NFV scenario.

The above modeling is for CPU intensive tasks, which is mainly taking CPU (computational) loads into consideration. In CloudSim energy consumption model, it mainly considers CPU loads. In our model, we extend to the CloudSim to capture memory, IO, network loads by adding the respective parameters and methods.

The loads can be predefined in workload configuration files. The workload configuration files can be divided as four types based on different resource type, including CPU workload file, memory workload file, disk storage workload file and bandwidth workload file. Each file is composed of many lines, each line represents a time interval and includes a 0-100 numeric value, which represents the load in this time interval. Like in the CPU load file, in line 1, the value is 40, which means in the first time interval, the CPU load is 40\%. By reading this file, the NFV task generator would generate the NFVlet with 40\% CPU loads in the first time interval. The NFVlet with more types of loads could also be generated in the same way.

The loads could also be generated with different distributions to control the number and length of tasks. Table I lists some possible compositions with distributions to generate loads:
\begin{table}
\footnotesize
\caption{compositions with distributions for loads generation}
\begin{center}
\begin{tabular}{|l|l|l|}
\hline Task length distribution& Task number & Application Type
\\\hline
\hline Uniform Distribution & 100 & CPU Intensive  \\
\hline Uniform Distribution & 100 & I/O Intensive  \\
\hline Uniform Distribution & 100 & Hybrid \\
\hline Normal Distribution units& 1,000 & CPU Intensive \\
\hline Normal Distribution & 1,000 & I/O Intensive \\
\hline Normal Distribution & 1,000 & Hybrid \\
\hline Poission Distribution & 10,000 & CPU Intensive  \\
\hline Poission Distribution & 10,000 & I/O Intensive \\
\hline Poission Distribution & 10,000 & Hybrid \\
\hline
\end{tabular} \\
\end{center}
\end{table}

After the loads are generated as NFVlets, NFVlets would be allocated to VM to form loads for VM, then to PM (Host), and finally forms the loads in DataCenter. Fig. 3 shows the load generation architecture of CloudSimNFV.

\begin{figure}[!ht]
\begin{center}
{\includegraphics [width=0.5\textwidth, height=1.5in,angle=-0] {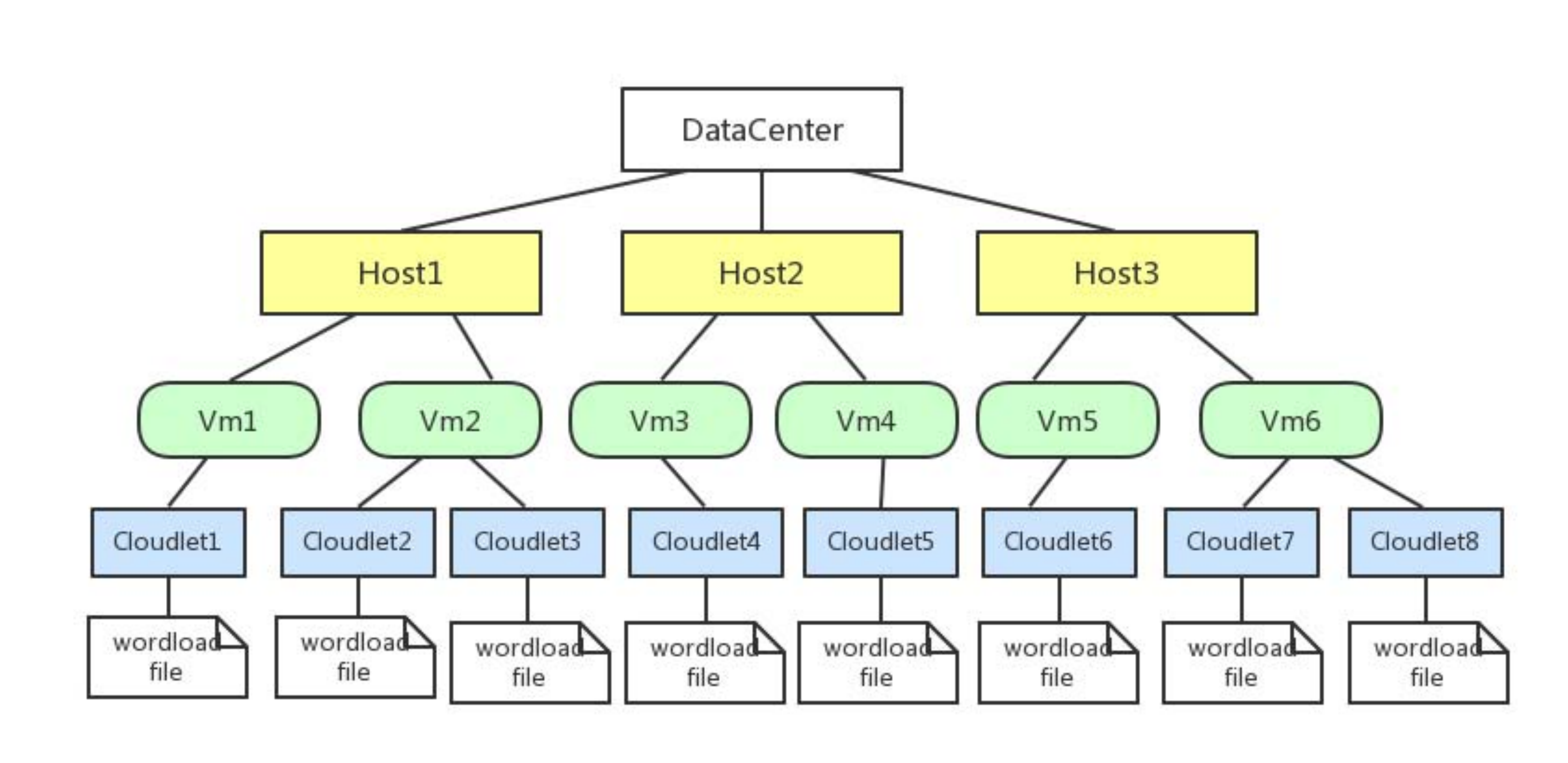}}
\caption{CloudSimNFV Load Generation Architecture}
\end{center}
\end{figure}

\subsection{User Interface}

\begin{figure}[!ht]
\begin{center}
{\includegraphics [width=0.5\textwidth, height=1.8in,angle=-0] {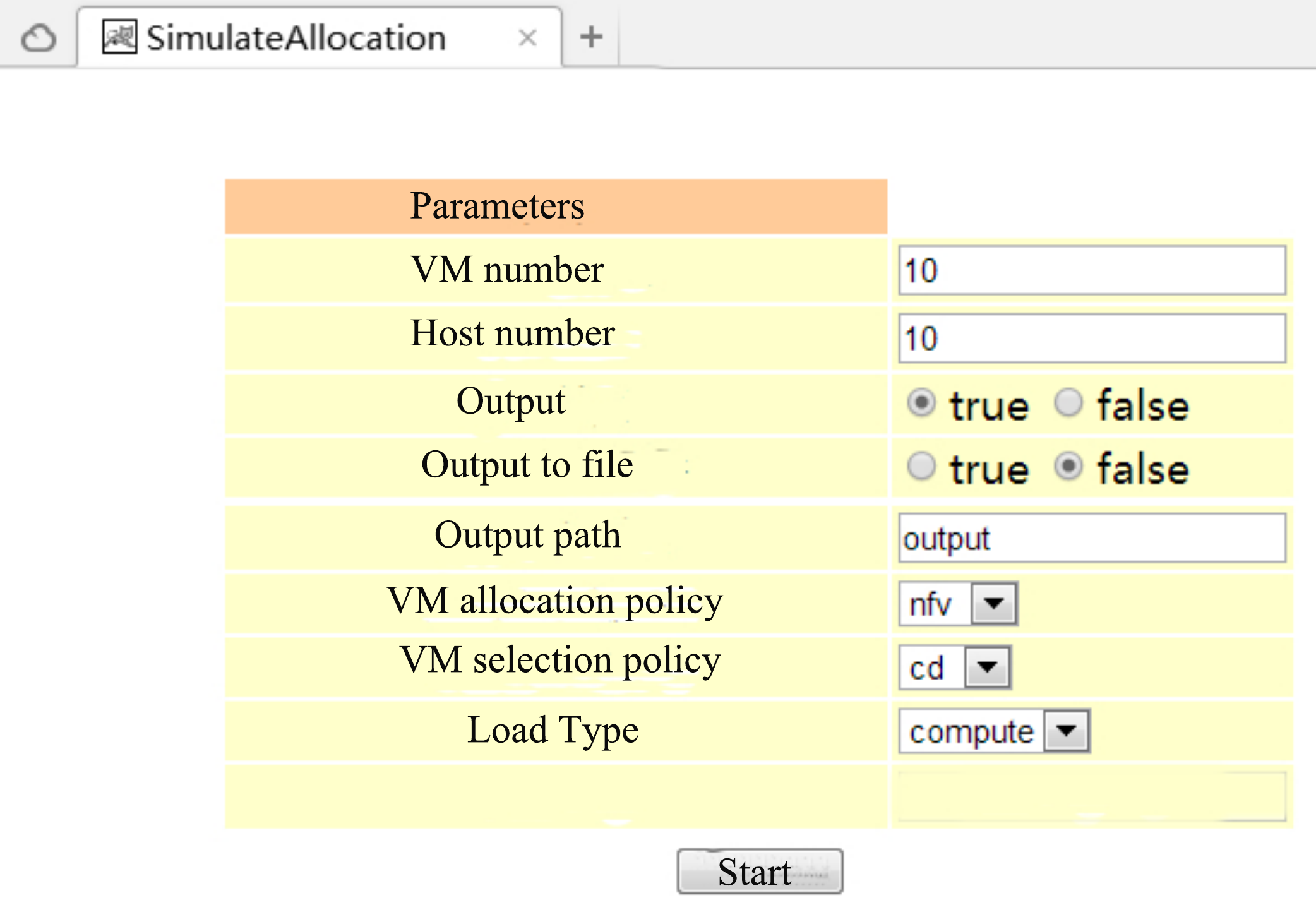}}
\caption{Simulation Configuration Interface}
\end{center}
\end{figure}

\begin{figure}[!ht]
\begin{center}
{\includegraphics [width=0.5\textwidth, height=1.5in,angle=-0] {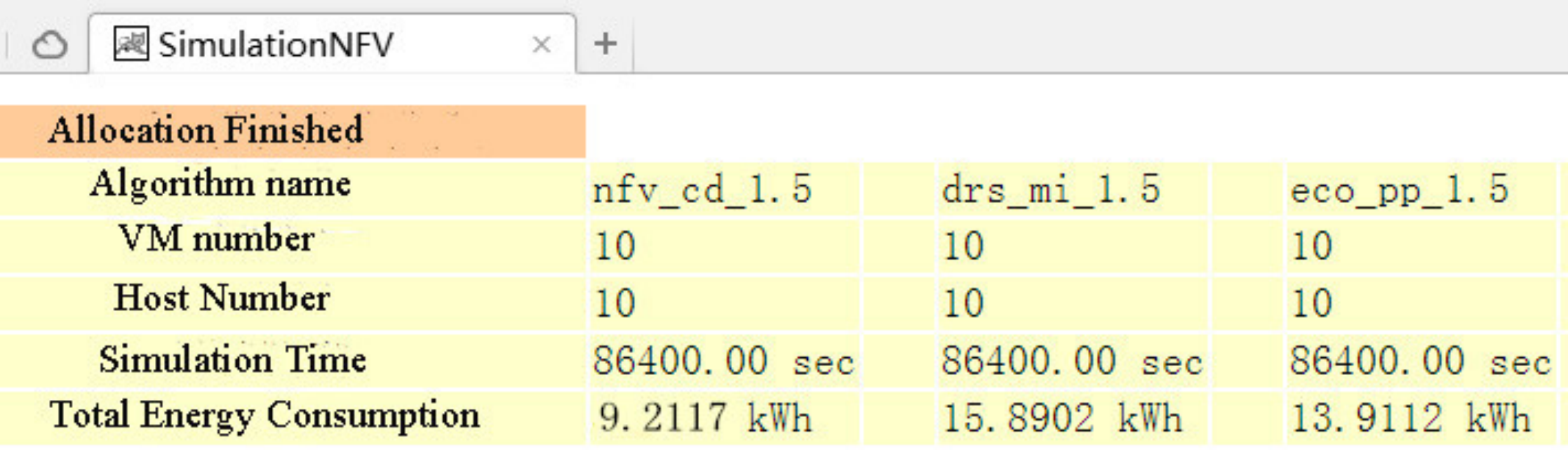}}
\caption{Simulation Results Interface}
\end{center}
\end{figure}

To simplify the user configuration for simulation, CloudSimNFV provides user portal to help user configure related parameters. Fig. 4 shows the portal page to set simulation parameters, including VM number, host number, output parameters, the algorithms for comparison (further description in Section V), load type for tests, and etc. Users can edit the parameters as they want, then clicking the Start Button to start simulation.

Fig. 5 demonstrates the corresponding simulation results after starting the configured simulation. The results show the basic simulation information and compared the energy consumption values.

\section{validation and Performance Evaluation}
\subsection{Validation}
Validation of CloudSimNFV is essential when it comes to persuasion of simulation. To fulfill this goal, we design and implement a set of experiments to validate by comparing CloudSimNFV and VMware under same workload and duration. VMware is a virtualization platform that enables to manage virtual machines on physical machine. VMware could produce more realistic data and has been equipped with energy efficiency management techniques and resource scheduling strategies. In our experiments, we setup environment that are composite of physical nodes that host VMware tool to manage virtual machine allocation and monitor energy consumption. Then our goal is to compare the energy consumption under different durations between hosts in CloudSimNFV and VMware, so that we can validate the accuracy of CloudSimNFV.

\subsubsection{Environment Setup}
To model the scenario that has loads fluctuation, we adopted a tool named stress-ng \cite{Stress} to generate load for the virtual machines in VMware to trace the energy consumption tendency. We also implemented the same load type in the way that described in earlier section IV.C. Under this environment, we setup three physical nodes (two nodes are with 12 cores and 48GB memory, another is with 8 cores and 4GB memory, one core capacity is equivalent to Intel Xeon CPU E5-2620 capacity), and each node can support several virtual machines (virtual machines are configured as one type of 4cores with 4GB or 2cores with 2GB, ). In addition, we created both physical machines and virtual machines with the same types in CloudSimNFV. Hence, the loads, virtual machines and physical machines have the corresponding elements both in realistic environment VMware and simulation tool CloudSimNFV.
\subsubsection{Validation Results}
\begin{figure}[!ht]
\begin{center}
{\includegraphics [width=0.5\textwidth, height=1.5in,angle=-0] {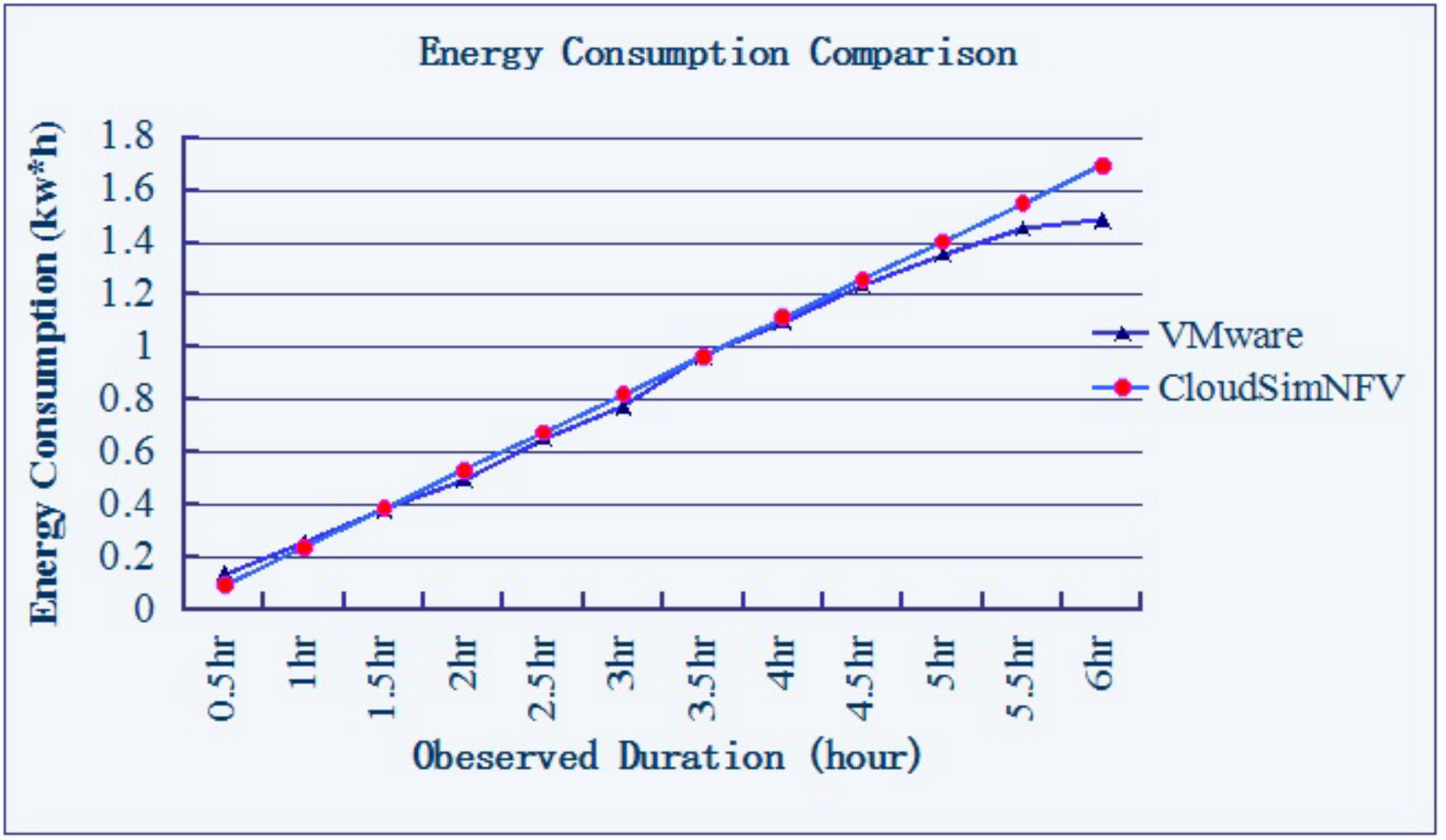}}
\caption{Simulation Results Interface}
\end{center}
\end{figure}

Fig. 6 demonstrates the obtained energy consumption in VMware and CloudSimNFV under the configured experiment in the earlier section. In this scenario, the total observed duration is 6 hours and we have varied observed periods, like $0-0.5hr$, $0-1hr$, \ldots, $0-6hr$ to trace the energy consumption fluctuation precisely. In this scenario, the differences between VMware and CloudSimNFV are always less than 5\%. This slight loss of accuracy is resulted from bandwidth energy consumption. We are still working on modeling the bandwidth energy consumption precisely considering data transferring.

\subsection{Energy Consumption Performance Evaluation}
In this paper, we adopt the Amazon EC2 configuration of VMs and PMs as shown in Table II and III. Note that one compute unit (CU) has equivalent CPU capacity of a 1.0-1.2 GHz 2007 Opteron or 2007 Xeon processor \cite{IEEEhowto:Amazon}.
\begin{table}
\footnotesize
\caption{8 types of virtual machines (VMs) in Amazon EC2}
\begin{center}
\begin{tabular}{|l|l|l|l|}
\hline Compute Units& Memory & Storage& VM Type
\\\hline
\hline 1 units & 1.7GB & 160GB & 1-1(1) \\
\hline 4 units & 7.5GB & 850GB & 1-2(2) \\
\hline 8 units & 15GB & 1690GB&1-3(3) \\
\hline 6.5 units& 17.1GB & 420GB &2-1(4)\\
\hline 13 units & 34.2GB & 850GB &2-2(5)\\
\hline 26 units & 68.4GB & 1690GB &2-3(6)\\
\hline 5 units & 1.7GB & 350GB &3-1(7) \\
\hline 20 units & 7GB & 1690GB &3-2(8)\\
\hline
\end{tabular} \\
\end{center}
\end{table}

\begin{table}
\caption{3 types of physical machines (PMs) suggested
}
\begin{center}
\footnotesize
\begin{tabular}{|l|l|l|l|}
\hline PM Pool Type& Compute Units & Memory& Storage
\\\hline
\hline Type 1 & 16 units & 30GB & 3380GB \\
\hline Type 2 & 52 units & 136GB & 3380GB \\
\hline Type 3 & 40 units & 14GB & 3380GB\\
\hline
\end{tabular} \\
\end{center}
\end{table}

In this section, we provide simulation results for comparing 3 different energy efficient scheduling algorithms for NFV scenario. The compared algorithms are as following:

\begin{table*}
\caption{Loads in One Day's Trace}
\begin{center}

\begin{tabular}{|l|l|l|l|l|l|l|l|l|l|l|}
\hline  & & \multicolumn{8}{| c |}{{Time Period (hour in one day)}} \\
\hline   \multicolumn{2}{| c |}{Load Type} & 0-2 & 2-6& 6-8 & 8-12 & 12-14 & 14-18&18-23 & 23-0\\
\hline
\hline \multirow{4}{2cm}{Computational Intensive Loads} & CPU Load & 30\% & 10\% & 30\% & 70\% & 60\% & 50\% & 90\% & 50\%  \\
& Memory Load & 30\% & 20\% & 30\% & 40\% & 40\% & 40\% & 50\% & 40\%  \\
 & Storage Load  & 20\% & 20\% & 20\% & 20\% & 20\% & 20\% & 20\% & 20\%  \\
 & Bandwidth Load & 20\% & 20\% & 20\% & 20\% & 20\% & 20\% & 20\% & 20\%  \\
\hline
\end{tabular} \\
\end{center}
\end{table*}

\begin{table*}
\caption{Energy Consumption and Migration Times Comparison}
\begin{center}
\footnotesize
\begin{tabular}{|l|l|l|l|l|l|}
\hline Test Case& Configuration & Comparison Index & DRS  & EcoCloud & NFV\\
\hline
\hline \multirow{2}{2cm}{Case 1} & \multirow{2}{2cm}{100HOSTS, 100VMS} & Energy Consumption (KW*h) &50.7217 & 45.2440 & 34.7679 \\
 &  & Migration Times &216 & 427 & 176 \\
\hline \multirow{2}{2cm}{Case 2} & \multirow{2}{2cm}{100HOSTS, 150VMS} & Energy Consumption (KW*h) &74.0505 & 66.1416 & 52.5940 \\
  &  & Migration Times&244 & 120 & 242 \\
\hline \multirow{2}{2cm}{Case 3} & \multirow{2}{2cm}{200HOSTS, 100VMS} & Energy Consumption (KW*h) & 48.2157 & 45.2463 & 35.0701\\
 &  & Migration Times& 298 & 240 & 312\\
\hline
\end{tabular} \\
\end{center}
\end{table*}

1. \emph{DRS (Dynamic Resource Scheduling) algorithm} \cite{Gulati}: In the initial allocation stage, the algorithm always allocates the VM to PM with the lowest load (typically the load means cpu utilization). DRS algorithm predefines a threshold that controls system imbalance level. During the migration stage, when the system imbalance level $g$ surpasses the predefined threshold, the migration algorithm would allocate VM to another PM that could decrease the $g$ value to be below threshold.

2. \emph{EcoCloud algorithm} \cite{Mastroianni}: The ecoCloud algorithm adopts a probabilistic function to select the most suitable PM to allocate VM. This algorithm would calculate a probabilistic value for each PM based PM's utilization $u$. Both utilization upper threshold $T_a$ and lower bound $T_b$ are predefined. The probability of PM to be allocated is calculated as $ f_a(u)=\frac{1}{M_p}u^p(T_a-u)$, $0 \leq u \leq T_a$, and $M_p= \frac{p^p}{(p+1)^{(p+1)}}T_a^{(p+1)}$, where $p$ can be set as $2, 3, 4$ , when utilization of PM surpasses the upper threshold or less than the lower bound, the migration algorithm could reallocate VM to the PM with highest probability according to the probabilistic function.

3. \emph{NFV algorithm}: The NFV algorithm also adopts a probabilistic function to select the most suitable PM to allocate VM. The probabilistic function is $F=\frac{-f(x; \alpha,\beta)}{3} + 1$, where $-f(x; \alpha,\beta)$ is Beta distribution, which equals to $\frac{x^{\alpha-1}(1-x)^{\beta-1}}{\int_0^1 u^a(1-u)^{\beta-1}du}$. The predefined upper and lower bound are also set, when utilization of PM is not between upper threshold and lower bound, VM migration would be triggered to make the PM utilization falls into the predefined interval. The migration process is also referring to the probabilistic function, in which the migrated VM would be reallocated to the PM with highest probability.

Table IV presents the computational type loads in one day's trace that we adopt for tests. This load trace is according to communication provider that the loads has peak (hour 8-12) and nadir (hour 0-2), which is quite suitable for NFV scenario.

The NFVlets are generated with the loads in Table IV. Then NFVlets are allocated to VMs and VMs are allocated to PM according the scheduling algorithms. We vary the host number and VM number, and compare the energy consumption and migration times in Table V. From the table, we can notice that under this scenario, NFV algorithm has the best energy-saving effects. We also obtain comparison results of the I/O intensive type loads, loads generated with other distributions (as indicated in Table I). With page limitation, we omit the detailed comparison for other load generation types.

\section{conclusions and Future Work}
As complementary of Software-Defined-Network, Network Function Virtualization aims to leverage standard IT virtualization technology to consolidate various network equipment types onto industry standard equipments. Network Function Virtualization can be applied to any data plane packet processing and control plane function in fixed and mobile network infrastructures.

Considering the complexity of realistic testbed, it's difficult to test all cases under real environment. Simulation tool could accelerate the related research on NFV by testing more cases and various situations. We propose a novel and repeatable simulation tool CloudSimNFV for NFV scenario based on CloudSim, aiming to improve the related work on NFV.

We describe our framework design and its components in detail. Energy consumption model and loads generation approaches are also discussed. Validation experiments illustrate the accuracy of CloudSimNFV. Performance evaluations show the energy efficient algorithms performance under NFV scenario.

We would also like to extend CloudSimNFV with more scheduling algorithms for more various scenarios in the future. In realistic data NFV data centers, constraints may exist between different servers. In the later versions of CloudSimNFV, the customized constraints would also be supported.


\section*{Acknowledgement}
This research is partially supported by China National Science Foundation (CNSF) with project ID 61450110440.


\begin{thebibliography}{99}

\bibitem{IEEEhowto:Amazon}
Amazon, Amazon Elastic Compute Cloud, http://aws.amazon.com/ec2/, 2013.
\bibitem{Basta}
Basta A, Kellerer W, Hoffmann M, et al. Applying NFV and SDN to LTE mobile core gateways, the functions placement problem[C] Proceedings of the 4th workshop on All things cellular: operations, applications, and  challenges. ACM, 2014: 33-38.
\bibitem{IEEEhowto:Bolla}
Bolla R, Lombardo C, Bruschi R, et al. DROPv2: energy efficiency through network function virtualization[J]. Network, IEEE, 2014, 28(2): 26-32.
\bibitem{Buyya}
Buyya R, Yeo C, Venugopal S, Broberg J, and Brandic I, Cloud
computing and emerging it platforms: Vision, hype, and reality for
delivering computing as the 5th utility, Future Generation computer
systems, vol. 25, no. 6, pp. 599¨C616, 2009.
\bibitem{IEEEhowto:Calheiros}
Calheiros R, Ranjan R,Beloglazov A, et al., CloudSim: A Toolkit for Modeling and Simulation of Cloud Computing Environments and Evaluation of Resource Provisioning Algorithms, Software: Practice and Experience, vol.41, no.1, pp.23-50, 2011
\bibitem{Carella}
Carella G, Corici M, Crosta P, et al. Cloudified IP Multimedia Subsystem (IMS) for Network Function Virtualization (NFV)-based architectures[C]//Computers and Communication (ISCC), 2014 IEEE Symposium on. IEEE, 2014: 1-6.
\bibitem{Chiosi}
Chiosi M, Clarke D, Willis P, et al. Network functions virtualisation introductory white paper[C]//SDN and OpenFlow World Congress. 2012.
\bibitem{Economou}
Economou D, Rivoire S, Kozyrakis C, et al. Full-system power analysis and modeling for server environments[C]. International Symposium on Computer Architecture-IEEE, 200
\bibitem{Guerout}
Gu¨¦rout T, Monteil T, Da Costa G, et al. Energy-aware simulation with DVFS[J]. Simulation Modelling Practice and Theory, 2013, 39: 76-91.
\bibitem{Gulati}
Gulati A, Shanmuganathan G, Holler A, Ahmad I, Cloud-scale resource management: challenges and techniques, VMware Technical Journal, 2011.
\bibitem{IEEEhowto: Kliazovich}
Kliazovich D, Bouvry P, Khan S, Greencloud: A packet-level simulator of energy-aware cloud computing data centers. IEEE Conference on GlobalTelecommunications, pp.1-5, 2010.
\bibitem{Mastroianni}
Mastroianni C, Meo M, Papuzzo G. Probabilistic consolidation of virtual machines in self-organizing cloud data centers[J]. Cloud Computing, IEEE Transactions on, 2013, 1(2): 215-228.
\bibitem{Nunez}
Nunez A, Vazquez-Poletti J, Caminero A, et al., iCanCloud: A Flexible and Scalable Cloud Infrastructure Simulator, Journal of Grid Computing 10:185,
C209, 2012

\bibitem{Riggio}
Riggio R, Rasheed T, Granelli F. Empower: A testbed for network function virtualization research and experimentation[C], Future Networks and Services (SDN4FNS), 2013 IEEE SDN for. IEEE, 2013: 1-5.
\bibitem{Soares}
Soares J, Dias M, Carapinha J, et al. Cloud4nfv: A platform for virtual network functions[C] Cloud Networking (CloudNet), 2014 IEEE 3rd International Conference on. IEEE, 2014: 288-293.
\bibitem{Stress}
Stress-NG--Tool to load and stress your Ubuntu system, http://www.ubuntugeek.com/stress-ng-tool-to-load-and-stress-your-ubuntu-system.html, 2015
\bibitem{Tian1}
Tian W, Xu M, Chen A, et al. Open-source simulators for Cloud computing: Comparative study and challenging issues[J]. Simulation Modelling Practice and Theory, 2015.
\bibitem{IEEEhowto:Tian2}
Tian W, Zhao Y, Xu M, et al. A toolkit for modeling and simulation of real-time virtual machine allocation in a cloud data center[J]. Automation Science and Engineering, IEEE Transactions on, 2015, 12(1): 153-161.

\bibitem{Xu}
Xu M, Tian W, Wang X, et al. FlexCloud: A Flexible and Extendible Simulator for Performance Evaluation of Virtual Machine Allocation[J]. arXiv preprint arXiv:1501.05789, 2015.
\bibitem{Udupi}
Yathiraj U, Debo D, Ramki, Increasing Infrastructure Efficiency via Optimized NFV Placement in OpenStack Clouds, OpenStack Atlanta Summit, 2014











\end{thebibliography}
\end{document}